\documentclass[twocolumn,twoside,slac_two]{revtex4}
\usepackage{graphicx}
\usepackage{fancyhdr}
\begin{document}

\title{Exploring the FRI/FRII radio dichotomy with the Fermi satellite}

%

\author{P. Grandi \& E. Torresi}
\affiliation{INAF/IASF Bologna, via Gobetti 101, 40129, Bologna, Italy}
\author{on behalf of the Fermi LAT collaboration}

\begin{abstract}
Misaligned Active Galactic Nuclei (MAGNs), i.e., radio galaxies and quasars with the jet not directly pointing at the observer, are a new class of GeV emitters.
In low power radio galaxies (i.e., FRIs), $\gamma$-rays are mainly produced in compact jet  regions, although in at least one case, Centaurus A, high energy photons from the radio lobes have been also observed.   The  first localization of the $\gamma$-ray dissipation zone in a high power radio galaxy (i.e., FRII) excludes major contributions from extended regions.  The study of  the FRII source  3C~111  indicates that $\gamma$-ray photons are produced in the jet.  The site, coincident with the radio core, is estimated to be at  a distance $\lesssim 0.3$ pc from the black hole.  Although the place where high energy photons are produced is probably similar in FRIs and FRIIs,  high power radio galaxies are rarer in the GeV sky.
Our  study of all the radio sources belonging to four complete radio catalogs (3CR, 3CRR, MS4, 2Jy) disfavors the idea that the paucity of  FRIIs is due to their larger distance (and therefore to their faintness) and supports  other possibilities, pointing to beaming/jet structural differences between  FRIs and FRIIs.

\end{abstract}

\maketitle

\thispagestyle{fancy}


\section{Introduction}
The LAT instrument \cite{lat} on--board the {\it Fermi} satellite has detected more than one thousand extragalactic sources in two years of survey \cite{2LAC}.
Unsurprisingly, the majority of these sources are blazars, which included BL Lacs (BLs) and Flat Spectrum Radio Quasars (FSRQs),  having the jet pointed  towards the observer. Due to Doppler boosting effects, the jet fluxes are amplified, making blazars the brightest sources in the extragalactic GeV sky.
Only a small part (3$\%$) of LAT detections have different counterparts. On the basis of the source of energy for the nonthermal particles, 
the non-blazar sources can be divided into two broad classes, i.e., Starbursts (SBs)  and AGNs.  In SBs, the acceleration of cosmic rays (protons and nuclei) is due to supernovae ejecta. The accelerated particles produce $\gamma$-ray photons interacting  with interstellar gas and radiation \cite{sb}. In AGNs, including Narrow Line Seyfert 1s \cite{fos}  and MAGNs, the particle accelerator is the black hole.  
MAGNs, the sub-class investigated here,  consist of radio galaxies (RGs) and Steep Spectrum Radio Quasars (SSRQs), i.e., Radio--Loud objects with steep radio spectra ($\alpha_r >$0.5) and/or possibly symmetrical extension in radio maps. RGs are divided into two radio morphological classes corresponding to edge-darkened (FRI) and edge-brightened (FRII) sources, with FRIIs being more powerful (P$_{178~MHz} > 10^{25}$ W Hz$^{-1}$ sr$^{-1}$). In FRIs, the jets are thought to decelerate and become sub-relativistic on scales of hundreds of pc to kpc, while the jets in FRIIs are at least moderately relativistic and supersonic from the core to the hot spots. FRIs and FRIIs are considered the parent population of BLs and FSRQs, respectively \cite{urry}. 
 \begin{figure}
\includegraphics[width=65mm]{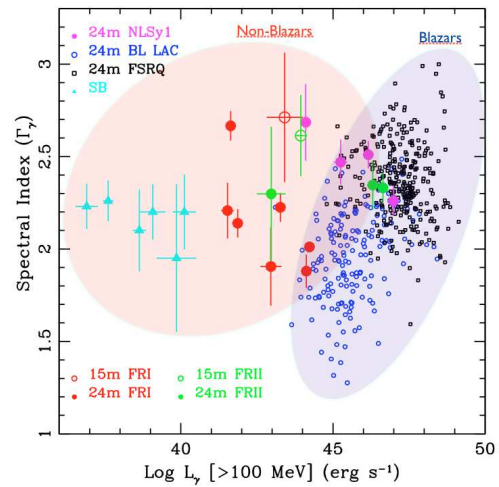}
\caption{Blazar and non-blazar sources after 24 months of {\it Fermi}-LAT survey. The spectral slopes are plotted as a function of the $\gamma$-ray luminosity (0.1-10 GeV). }
\label{fig1}
\end{figure}

The first sample of MAGNs was collected in 15 months of data \cite{magn} and consists of 11 sources, characterized by low $\gamma$-ray fluxes (F$_{\rm > 0.1 GeV} \sim$10$^{-8}$ phot~cm$^{-2}$~s$^{-1}$) and generally steep spectral slopes ($\Gamma >$2.4).   In spite of their small number, the non-blazar $\gamma$-ray emitters are extremely appealing, as they  offer  powerful physical evidence for investigating high energy phenomena.
The study of the Spectral Energy Distribution (SED) of single FRI radio galaxies has revealed, for example,  that the assumption of a pure, one-zone homogeneous, Synchrotron Self-Compton (SSC) emission region is inadequate. The one-zone SSC model can fit the data only assuming slow jets characterized by a Lorentz factor $\Gamma_{FRI}$=2-3\footnote{The Doppler factor of a moving source is defined as $\delta=[\Gamma(1-\beta cos\theta)]^{-1}$, where $\beta$ is the bulk velocity, $\Gamma$ the Lorentz factor and $\theta$ is the jet inclination angle} significantly smaller than those required in BLs ($\Gamma_{BL}\ge 10$).
This appears in conflict with the unified models that assume FRI radio galaxies to be  the parent population of BLs, i.e., the same type of object seen at larger inclination angles. Different solutions have been proposed to  explain this inconsistency \cite{greco, ghisellini, bo}.  The new models relax the idea of one homogeneous zone and assume different interacting regions in the jet.

A very intriguing aspect related to the MAGN sample is the predominance of FRI RGs. FRIIs appear to be more elusive objects, as also attested by a study searching for $\gamma$--ray counterparts of Broad Line Radio Galaxies (BLRGs) \cite{kataoka}.  
The simplest interpretation is that FRIIs are locally rarer and therefore more difficult to detect because of their fainter fluxes.  
Here we investigate this possibility following two different strategies:  i)  the identification of  the $\gamma$-ray emitting regions in RGs,  with particular attention to 
3C~111, the  only FRII  in addition to  Pictor A\cite{brown} , having  a secure {\it Fermi}-LAT counterpart; ii)  a comparative study of the FRI and FRII radio and $\gamma$-ray  properties of four complete samples of radio sources.

\section{Where is the source of the high energy photons in RGs?}

Variability studies can provide important information on  the size of the emitting region. Using causality arguments, it is possible to constrain the emitting region dimension assuming 
 $R\le c\Delta t\delta/(1+z)$,  where $\Delta t$ is the time variability and $\delta$ the Doppler factor.  
This approach does not  allow a direct localization of  the emitting region, unless the dissipation  is assumed to take  place in the entire cross section of the jet.  In this case, if the half-opening  angle of the jet  ($\varphi$) is known,  a roughly estimation of  
the gap ($d$) between the base of the jet and the gamma-ray source is  $d=R/sin\varphi$. 
Recent results, based on TeV variability studies of FSRQs \cite{aleski}--\cite{tavpks}, have made evident that this assumption is  probably not correct. 
 The observed Very High Energy (VHE)  flares, occurring on time scale of minutes,  constrain the emitting region within the Broad Line Region. On the other hand, this zone, rich in UV and optical radiation, is expected to be opaque to photons with energies   $E_{\gamma\gamma}\sim 60 (\frac{10^{15}}{\nu})$  GeV \cite{aleski} because of  pair production. This discrepancy, having a strong impact on theoretical models, has  triggered variability studies.
 Multiwavelength  techniques seem to be a particularly  powerful instrument to address this point. The blazar analysis of outbursts at different frequencies has revealed their stochastic nature and the difficulty in defining a unique dissipation zone. The events can occur in radio cores as well as in knots along the jet \cite{agudo}--\cite{marscher}.
 Constraining the dimension and the position of the high energy emitting region in MAGNs is a more difficult task. These sources are indeed  faint at  GeV energies and rare in the TeV sky.  Incidentally, we note that only three FRI RGs, i.e., M87 \cite{m87tev}, NGC~1275 \cite{m1275},
 and Centaurus A \cite{cenatev}--\cite{ic310}\footnote{Another RG observed at TeV energies is IC~310. It is not part of the MAGN sample.} have been discovered  by Cherenkov telescopes up to now.
Among FRIs, NGC~1275 is the only source showing significant $\gamma$-ray flux changes  on time scales of months \cite{magn}--\cite{1275k}--\cite{brown} and 
M87 is the only RG  exhibiting TeV flare on time scale of days.  In agreement with the blazar results,  
ten years  monitoring M87 seems to indicate multiple emitting regions of VHE photons. TeV  flares appeared to be  simultaneous with a strong X-ray burst observed in a jet knot in 2005, to a strong increase (and successive knot ejection) of the radio core in 2008, and to a nuclear X-ray increase in 2011 \cite{abramowski}.  The observed variability clearly supports  the compactness of the GeV/TeV sources in FRIs, although the discovery of  $\gamma$-ray counterparts of the radio lobes in Centaurus A \cite{cena} slightly confuses the picture.
Conversely, the dimension/location of the $\gamma$-ray regions (jet, hot spots lobes?)  in FRIIs is completely unknown. The first result,  presented at this meeting for the first time and discussed in detail by \cite{3C111}, concerns 3C~111.
\begin{figure}[th]
\includegraphics[width=80mm]{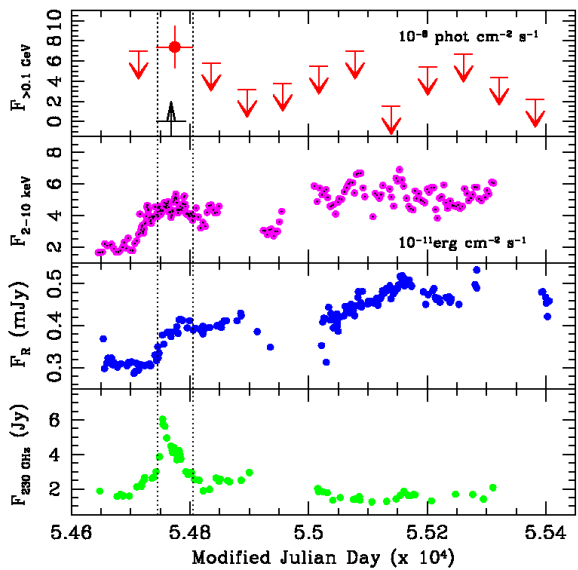}
\caption{mm (230 GHz), optical (R), X-ray (2.4-10 keV) and $\gamma$-ray (0.1-100 GeV) light curves of 3C~111. The mm-optical-X-ray data are from \cite{chatterjee}. The dotted lines limit the flare event (occurring between 2008 October 4 and December 4). The black arrow indicates the time when the radio knot was ejected by the core.}
\label{fig2}
\end{figure}
3C~111(z=0.0491) is a BLRG  exhibiting radio lobes, a strong core and a knotty jet. It is  a source extensively monitored from radio to X-rays by the blazar research group at  Boston University (BU)\footnote{www.bu.edu/blazars/}. 3C~111 is listed in the first\cite{1Lac} but not in the second LAT Catalog of AGN\cite{2LAC}. Indeed, in the first two years of the survey, the source appeared  in a high state only at the end of 2008 and below the LAT sensitivity threshold for the rest of the time. 
An on-going radio to X-ray monitoring campaign\cite{chatterjee} clearly showed that 3C~111 was detected by LAT exactly when the mm--to--X--ray fluxes were increasing (Figure 2), unambiguously attesting  to co-spatiality of the events. Chatterjee et al. \cite{chatterjee}  noted that the X-ray variations are strongly correlated with those at  optical frequencies and that a pronounced minimum in the 2-10 keV light curve comes before the ejection of a new radio knot from the core. 
As most of the optical and X-ray photons are probably produced in a disk-corona (or ADAF-disk) system, they interpreted the observed X-ray dips as signatures  of  possible jet/accretion-flow perturbations.\\
Looking at Figure 2, it appears evident that the $\gamma$-ray outburst occurred after a strong decrease of the X-ray flux and during the ejection of a new radio blob. 
The Very Long Baseline Array (VLBA) images of 3C~111 at three different epochs of 2008 is shown in Figure 3.  The radio core is weak in  August,  very bright in November and split into two components in December. This jet modification roughly follows the $\gamma$-ray flux evolution, revealing a direct connection between  the $\gamma$-ray outburst  and the ejection of a new radio knot.  This unambiguously localizes the  $\gamma$--ray source in the radio core.
\begin{figure}[t]
\includegraphics[height=50mm,width=82mm]{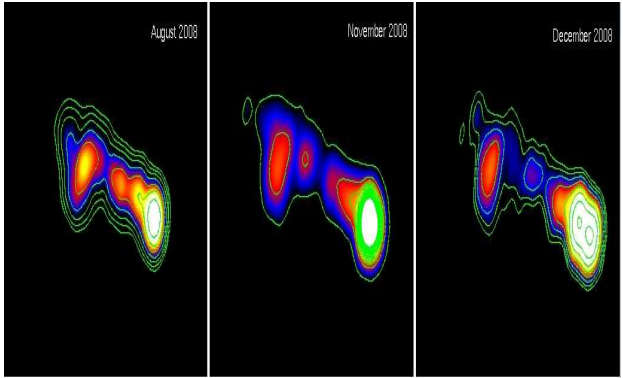}
\caption{Sequences of the 43 GHz VLBA images of 3C~111 before and after the $\gamma$-ray  flare.  These observations were performed by the BU group.}
\label{fig3}
\end{figure}
As discussed in detail by \cite{3C111}, a  feature  is seen in the VLBA maps  at 0.1 mas ($\sim$ 0.3 pc) from the core,  just on the opposite side of the ejected bright knot. If this is the counter-jet, the black hole necessarily lies in between (and therefore at less than 0.1 mas from the radio core).  Considering that the  $\gamma$-ray source is within the radio core and that the VLBA spatial resolution  is also $\sim0.1$ mas,  we can conclude that  the distance of the $\gamma$-ray dissipation region from the central engine is  $\sim 0.3$ pc or slightly larger depending on its position with respect to the counter-jet. Following the prescriptions of \cite{GT}\footnote{The distance of the Broad Line Region and the torus can be directly related to the disk luminosity. The optical luminosity L$_V=6 \times 10^{43}$ erg $s^{-1}$ of 3C~111 (provided by \cite{chatterjee} and rescaled to our cosmology) was converted to the disk luminosity assuming  a bolometric correction of $\sim 10$ \cite{Elvis}.}, we estimated the distance of the Broad Line Region and the torus from the black hole to be 0.02 and 0.6 pc, respectively.  The compact  $\gamma$-ray source of the  FRII RG 3C~111, being  at $\sim 0.3$ pc from the central engine,  is then necessarily confined within the torus.

\section{Why does the LAT miss FRIIs?}
In order to answer this question we consider the 3CR \cite{bennett}--\cite{3cr}, 3CRR \cite{3crr}, Molonglo (MS4) \cite{ms4a}--\cite{ms4b} and 2Jy \cite{2jy}--\cite{morganti} radio catalogs.
 We intentionally  chose the low (178 MHz, 408 MHz) radio frequency 3CR, 3CRR and MS4 catalogs because they preferentially select radio sources characterized by steep--spectrum synchrotron emission from extended lobes.  To directly compare  MAGN and blazar properties, we took also into account the 2Jy sample{\footnote{http://2jy.extragalactic.info/2Jy$_{-}$home$_{-}$page.html}.
This survey, collecting  southern galaxies with flux larger than 2 Jy at 2.7 GHz, includes several compact sources.
The total sample obtained by combining all the radio catalogs (3C-MS-2Jy sample)  was then  cross--correlated with the second AGN LAT catalog \cite{2LAC}.
\begin{figure}
\includegraphics[width=63mm]{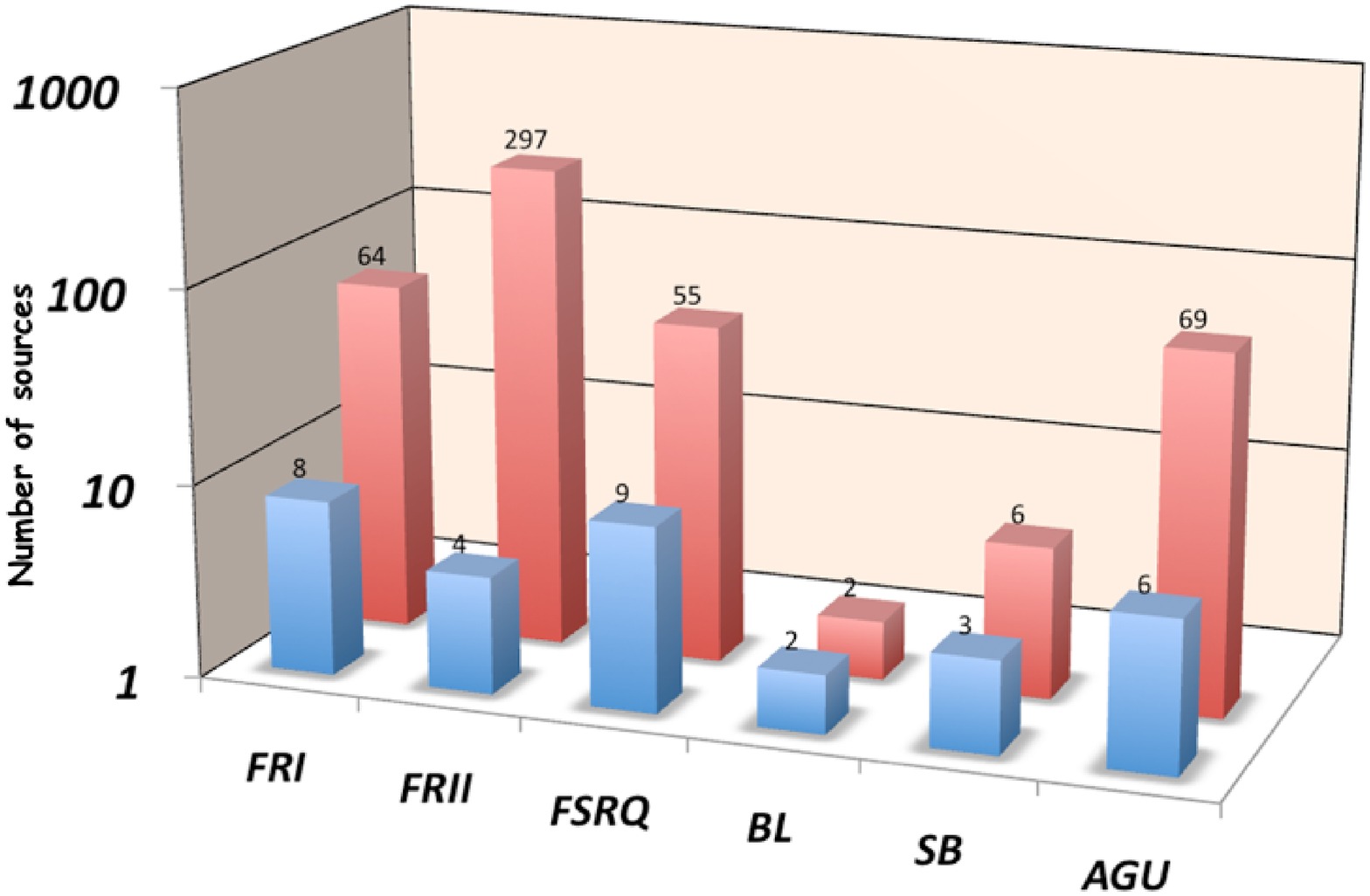}
\includegraphics[width=63mm]{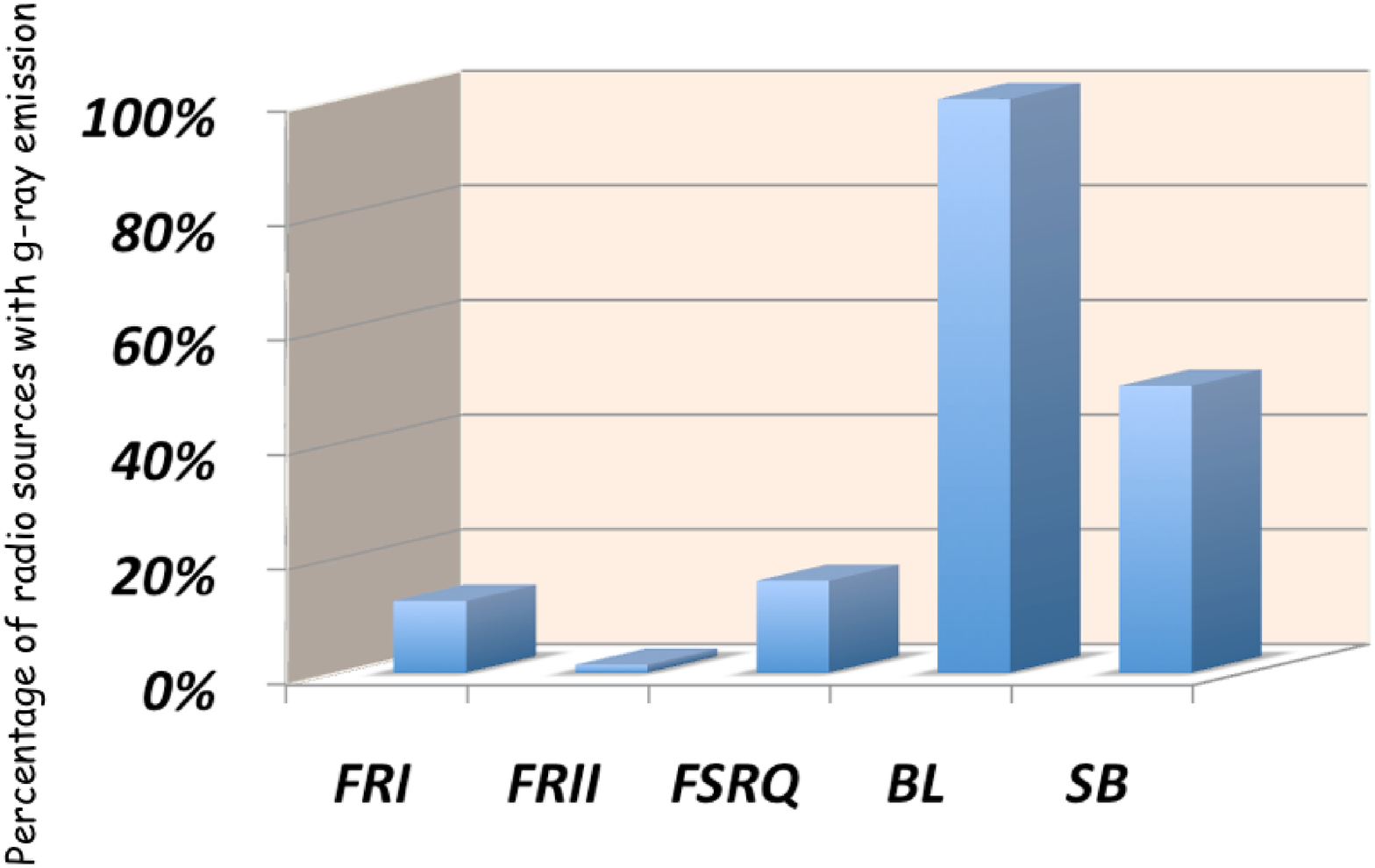}
\caption{{\it Upper panel} -- The red histogram indicates the number of sources in the 3CR-MS4-2Jy sample classified as FRI, FRII, FSRQ, BL, SB.  The AGU class collects objects with unknown classification.
The blue histogram shows how many objects for each class have a LAT counterpart. {\it Lower panel} -- LAT detection rate for each class of the 3CR-MS4-2Jy sample.}
\label{fig4}
\end{figure}
Figure 4 ({\it upper panel}) shows the demography of the combined 3C-MS4-2Jy sample (red histogram) together with  the LAT associations for each class of objects (blue histogram). 
FRIIs represent the most populous class and furthermore has the lowest number of LAT detections. 
This is even more evident in Figure 4 ({\it lower panel}) where the percentages of sources with a $\gamma$-ray counterpart are reported for each group (FRI, FRII, FSRQs, BLs, SBs and AGUs, i.e., AGNs with unknown classification).
BLs have the highest probability to be detected, directly followed by SBs. FSRQs and FRIs have similar detection rates. FRIIs, the most numerous objects in the 3C-MS4-2Jy sample, are rare with only 1$\%$ of them observed in the $\gamma$--ray sky. Note that similar detection rates are obtained considering each radio catalog separately.
\begin{figure}[b]
\includegraphics[width=48mm]{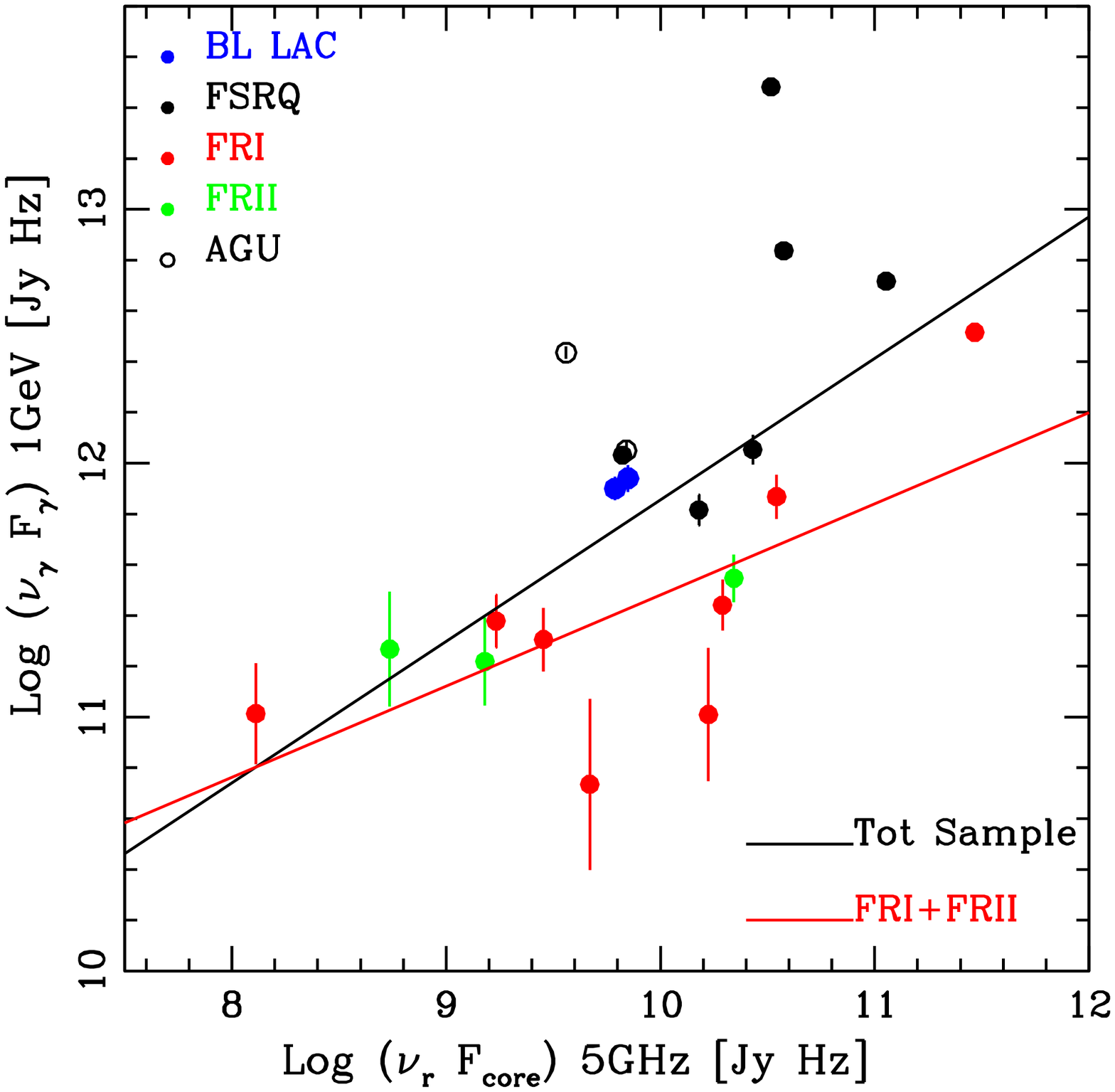}
\includegraphics[width=48mm]{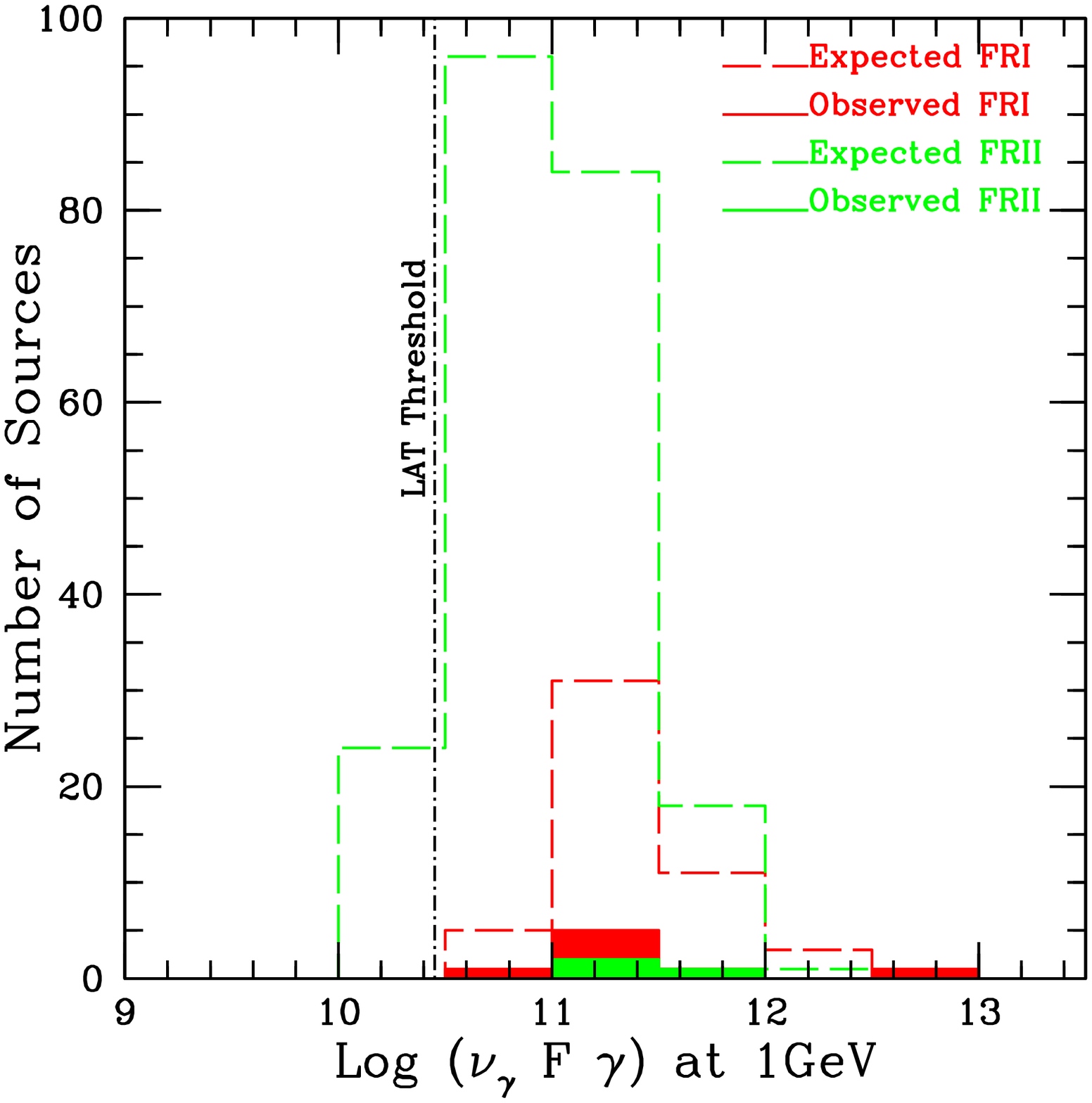}
\caption{{\it Upper panel} -- Radio-Gamma correlation of the 3C-MS4-2Jy  objects  with a $\gamma$-ray  association (black line). A positive  (flatter) trend is also present when only FRI and FRII sources are
  taken into account (red line).  {\it Lower panel} --  Histogram of the predicted fluxes for all the FRI (red) and FRII (green) radio sources belonging to the 3C-MS4-2Jy sample (dashed lines). For comparison, the real observed MAGNs are also shown. Many FRIIs are expected to be bright enough to be detected by {\it Fermi}}
\label{fig5}
\end{figure}
The simplest interpretation of this result is that {\it Fermi}  cannot detect FRIIs because they are far away and, therefore, not bright enough to be detected.
To verify this hypothesis, we attempted to estimate the  $\gamma$-ray fluxes of the MAGNs for which no GeV association
was found, taking advantage of the correlation between the radio core and GeV fluxes noted by some authors \cite{ackermann}--\cite{ghirlanda}.
We plotted the $\gamma$-ray flux (at 1 GeV) of the 3C-MS4-2Jy sources with LAT detections as a function of the radio core flux at 5 GHz, 
searching for a correlation. Note that,  we implicitly  assume that the high energy emission mainly  occurs in the jet (as supported by the observations - see previous section).  
As shown in Figure 5 ({\it upper panel}) and statistically demonstrated
in Table I, a positive trend is clearly present.
The probability (P$_r$) that the observed correlation
occurred by chance is less than 1$\%$   for  the 
total sample and is less than 3$\%$
when only FRIs and FRIIs are taken into account.
Knowning the radio core flux,  the expected GeV flux of each source of the 3C-MS4-2Jy total 
sample can be then estimated 
using the linear  relation between the radio core (at 5 GHz) and gamma-ray  fluxes (at 1 GeV)
(see Table I).
\begin{table}[h]
\begin{center}
\caption{\small: Radio core-$\gamma$-ray linear correlation:
$Log(\nu_{\gamma} F_{\gamma}) = {\bf a} + {\bf b} \times  Log(\nu_{r} F_{core})$ 
\\ {\bf r} is the Spearman coefficient of correlation and ${\bf P_r}$  
the relative probability that the correlation occurs by chance.}
\begin{tabular}{ccccc}
\hline 
\textbf{Sample} & \textbf{a} & \textbf{b} & \textbf{r} & \textbf{${\bf P_r}$}\\
\hline 
Total     & 6.2(1.6) & 0.6 (0.2)& 0.65& $<0.01$ \\
 MAGN&   7.9(1.2) & 0.4 (0.1) & 0.66& $<0.03$\\
\hline
\end{tabular}
\label{t1}
\end{center}
\end{table}

In Figure 5 ({\it lower
panel}), the histogram of the predicted (dashed lines)
and observed (solid lines)  $\gamma$-ray fluxes for both FRI
(red lines) and FRII (green lines) classes are shown. 
Here the estimated fluxes at 1 GeV are based on the linear regression
of the MAGNs alone. Similar results are 
obtained even when the blazars (plus AGUs) with LAT counterparts
are considered in fitting the radio-gamma data.\\
It is evident that, the number of  FRIs and FRIIs expected to have a gamma-ray flux
above the LAT sensitivity threshold (dotted vertical
line in Figure 5 {\it lower  panel}) is considerably larger that those actually detected. This could suggest 
that non-blazar radio sources are preferentially
detected during periods of strong jet activity. In any case, the gap between 
 the FRIIs potentially detectable  by {\it Fermi} and those really observed  is impressive and definitely larger than that of FRI class.
The idea  that distance/faintness effects can explain
the scarcity of powerful edge-brightened radio sources does not seem to be supported by the data.
Other interpretations need to be explored.\\
It is possible that the jet activity is more often triggered in FRIs than in FRIIs. 
Spending more time in a high state, low power radio galaxies could be more easily caught by the LAT.
Alternatively (but not necessarily in  contrast with the previous suggestion) 
the assumption of similar flux boosting
factors at low and high frequencies for both FRIs
and FRIIs could be too simple \cite{dermer}.
For example, if the emission
is due to Compton scattering of external photons in
the jet (EC) the Doppler boosting is stronger and the
beaming cone narrower than in the case of SSC processes (see Fig. 6). If the EC and SSC mechanisms
dominate the high energy emission of FRIIs and FRIs, respectively \cite{torresi},  a beaming difference could explain the lower
$\gamma$-ray detection rate of  powerful edge-brightened
radio galaxies\footnote{It is known that FRIIs are generally richer in seed photons
(coming from accretion disks, Broad Line Regions, tori and 
Narrow Line Regions) while  FRIs seem to be characterized by a poorer
environment.}.
 \begin{figure}
\includegraphics[width=45mm]{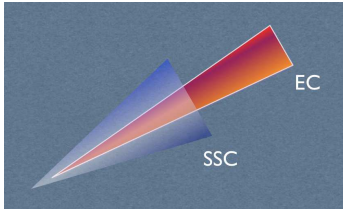}
\caption{The Doppler boosting is stronger and the
beaming cone narrower if the high energy emission is due to Compton scattering of jet environment photons (EC process) rather than 
to Compton scattering of synchrotron photons produced in the jet  (SSC process).}
\label{fig6}
\end{figure}
\begin{figure}
\includegraphics[width=43mm]{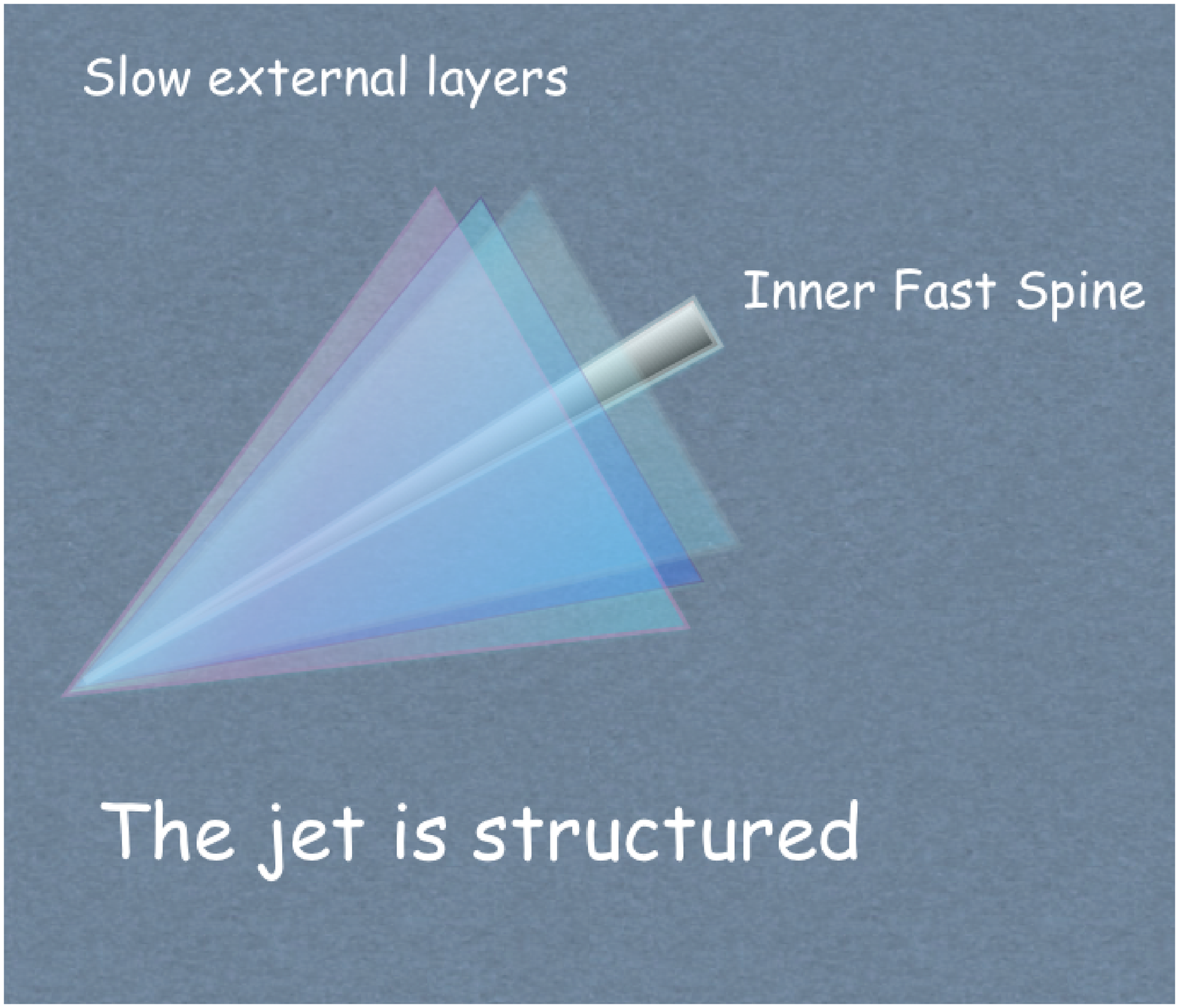}
\includegraphics[width=43mm]{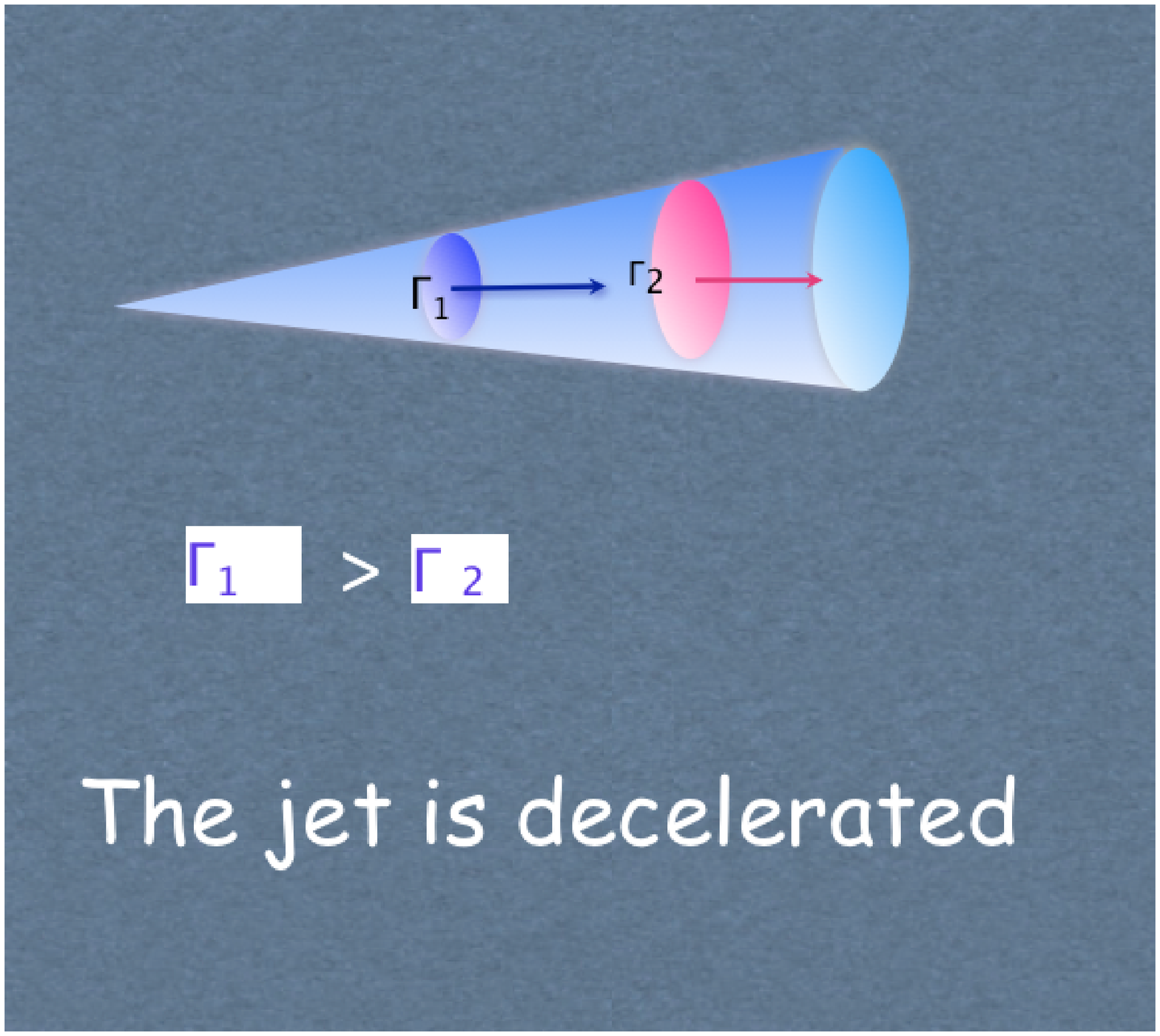}
\includegraphics[width=43mm]{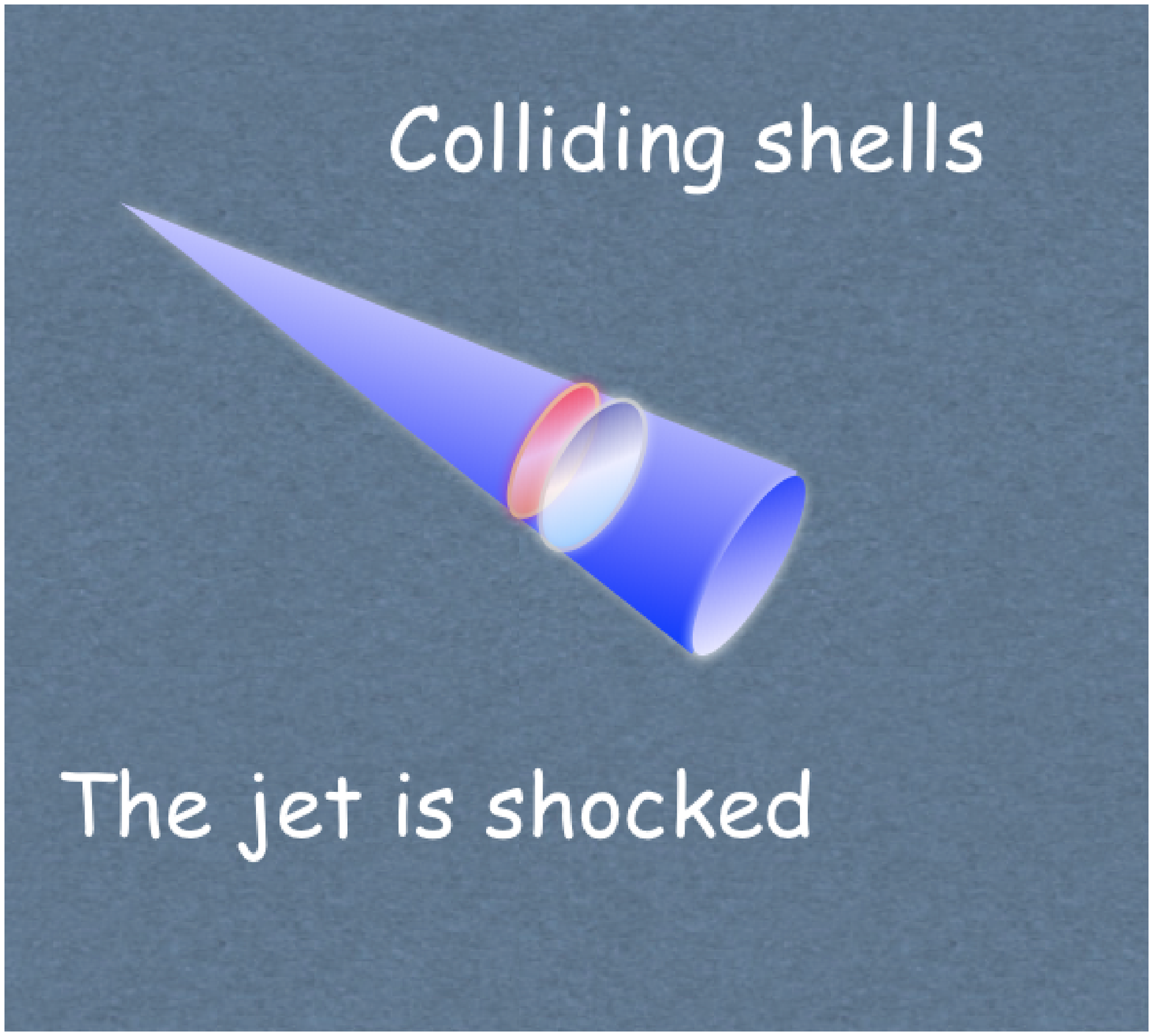}
\caption{A sketch of the recent models proposed for a jet.  The one-zone homogeneous emission region is inadequate to interpret the observations that require
 a more  complex jet.  The jet could be structured with an inner fast region surrounded by slower layers \cite{ghisellini} ({\it upper panel}), could be  decelerating \cite {greco} ({\it middle panel}) or  could contain colliding relativistic shells \cite{bo} ({\it lower panel}). }
\label{fig7}
\end{figure}
\noindent
Another possibility is  that one--zone homogeneous SSC/EC models are an oversimplified interpretation, as suggested by 
 the study of the FRI SEDs of  NGC~1275 \cite{1275}, M87 \cite{m87} and NGC~6251 \cite{giulia}.
New models have been proposed in the last years, like, for example, 
the decelerating jet flow \cite{greco},  the spine-layer jet \cite{ghisellini} and the colliding shell \cite{bo} models (Figure 7), 
all  assuming a structured jet with different regions at different velocities. 
Within this context, the paucity of FRIIs with a LAT counterpart, could be  ascribed  to
a less prominent (or absent) external layers and/or less efficient deceleration processes \cite{paola2011}.
Incidentally,  we note that the recent fast variability observed in  blazars  at TeV energies (see previous Section) has further weakened the one-zone homogeneous jet picture  \cite{tavpks}--\cite{dermerlott}.

\section{Conclusions}
The $\gamma$-ray study of MAGNs, a new class of GeV emitters discovered by the {\it Fermi}--LAT, is of primary importance in understanding the complexity of FRI and FRII jets. Indeed in RGs, high energy photons  are mainly produced in the jet, although  the {\it Fermi} observation of the radio lobes of Centaurus A indicates  
a possible contribution from extended kpc structures.  
Flux variability observed at high energies in FRI RGs (NGC~1275 and M87) supports the idea that the emitting regions are compact.  
The first localization of the $\gamma$--ray source in a FRII RG, i.e., 3C~111,  excludes a significant GeV contribution from radio lobes and/or hotspots.
The dissipation region, at least during a flare,  is within the radio core of the jet at a distance $<0.3$ pc from the black hole.\\
Although the place where high energy photons are produced  is similar in both FRIs and FRIIs,  high power radio galaxies are rarer in the GeV sky.
We confirmed the paucity of FRIIs, analyzing 4 complete catalogs of radio sources.
Although the majority of the objects  belongs to the FRII  class, only $1\%$  of them has a $\gamma$-ray counterpart.  Assuming as plausible the link between the 
radio core and the $\gamma$-ray fluxes, it is improbable that FRIIs are missed  by {\it Fermi}, because 
locally rarer and therefore fainter. We suggest other possible interpretations.  If radio galaxies are preferentially observed during a flare, it could be difficult 
to catch FRIIs in case  they spend most of the time in quiescent phase.  The FRII $\gamma$-ray elusiveness  could also reflect  intrinsic jet differences.   If the high energy emission is mainly due to EC in FRIIs   and to SSC  in FRIs \cite{torresi},  the Doppler boosting is expected to be  stronger and the beaming cone narrower in more powerful RGs.  
Finally,  FRII jets could be less structured, likely poorer  of shocked regions,  and dominated by a fast spine, while in FRIs,  the external layers could be more  prominent  and/or the jet could be more efficiently  decelerated.

\bigskip 
\begin{acknowledgments}
{\scriptsize The authors acknowledge the Boston University blazar research group, in particular S. G. Jorstad, A. P. Marscher and R. Chatterjee:
the study of 3C~111 was possible thanks to their seminal papers and data generously made available to the scientific community.
P.G. would like to thank  C. Dermer for a thorough reading of the manuscript and precious comments.
The \textit{Fermi} LAT Collaboration acknowledges generous ongoing support
from a number of agencies and institutes that have supported both the
development and the operation of the LAT as well as scientific data analysis.
These include the National Aeronautics and Space Administration and the
Department of Energy in the United States, the Commissariat \`a l'Energie Atomique
and the Centre National de la Recherche Scientifique / Institut National de Physique
Nucl\'eaire et de Physique des Particules in France, the Agenzia Spaziale Italiana
and the Istituto Nazionale di Fisica Nucleare in Italy, the Ministry of Education,
Culture, Sports, Science and Technology (MEXT), High Energy Accelerator Research
Organization (KEK) and Japan Aerospace Exploration Agency (JAXA) in Japan, and
the K.~A.~Wallenberg Foundation, the Swedish Research Council and the
Swedish National Space Board in Sweden. Additional support for science analysis during the operations phase is gratefully
acknowledged from the Istituto Nazionale di Astrofisica in Italy and the Centre National d'\'Etudes Spatiales in France}
\end{acknowledgments}


\bigskip \begin{thebibliography}{99} 

\bibitem{lat}{W.B. Atwood, et al. 2009, ApJ, 697, 1071}

\bibitem{2LAC} {M. Ackermann, et al.  2011, ApJ submitted, arXiv:1108.1420} 

\bibitem{sb}{A.A. Abdo et al.  2010b, ApJ, 709, L151}

\bibitem{fos}{A.A. Abdo et al.  2009, ApJ, 699, 976}

\bibitem{urry}{C.M. Urry, \& P. Padovani 1995, PASPJ, 107, 803}

\bibitem{magn} {A.A. Abdo et al.  2010c,  ApJ, 720, 912} (MAGN)

\bibitem{greco}{M. Georganopoulos, \& D. Kazanas  2003, ApJ, 589, L5}

\bibitem{ghisellini}{G. Ghisellini, F. Tavecchio, M. Chiaberge  2005, A\&A, 432, 401}

\bibitem{bo}{ M. B\"{o}ttcher and C. D. Dermer  2010,  ApJ, 711, 445}

\bibitem{kataoka}{J. Kataoka, et al. 2011 ApJ, 740, 29}

\bibitem{brown}{A. M. Brown, J. Adams  2011, MNRAS, 413, 2785}

\bibitem{aleski}{ J. Aleksic, et al. 2011, ApJL, 730, 8}

\bibitem{tavpks}{F. Tavecchio, J.  Becerra-Gonzalez, G. Ghisellini, et al.  2011, A\&A,  534, 86}

\bibitem{agudo}{I. Agudo, S.G. Jorstad, A.P. Marscher, et al.  2011, ApJ, 726, L13}

\bibitem{marscher}{A. P. Marscher, S.G. Jorstad, V.M. Larionov, et al.  2010, ApJ, 710, L126}

\bibitem{m87tev} {F. A.  Aharonian, et al. 2003, A$\&$A, 403, L1}

\bibitem{m1275} {J. Aleksic, et al.  2012,  A$\&$A in press, arXiv:1112.3917}

\bibitem{cenatev}{F. A.  Aharonian,  et al.  2009,  ApJL , 695, 40}

\bibitem{ic310}{J. Aleksic,  et al. 2010,  ApJ, 723, 207}

\bibitem{1275k}{J. Kataoka, et al.  2010, ApJ, 715, 554}


\bibitem{abramowski}{A. Abramowski,  et al.  2012, ApJ, 746, 151}

\bibitem{cena}{A. A.  Abdo et al.  2010d, Science, 328, 725}


\bibitem{3C111}{P. Grandi, E. Torresi, C. Stanghellini  2012, ApJL, 751, 3 }

\bibitem{1Lac}{A.  A. Abdo et al.  2010a, ApJS, 188, 405} (1 LAC)


\bibitem{chatterjee}{R. Chatterjee, A. P. Marscher, S. G. Jorstad, et al.  2011, ApJ, 734, 43}

\bibitem{GT}{G. Ghisellini $\&$ F. Tavecchio 2008, MNRAS, 387, 1669}

\bibitem{Elvis} {M. Elvis, et al. 1994 ApJS, 95, 1}

\bibitem{bennett}{A. S. Bennett 1962, MNRAS, 125, 75}

\bibitem{3cr}{H. Spinrad, et al. 1985, PASP, 97, 932}

\bibitem{3crr}{R. A. Laing, J.M. Riley, \& M.S. Longair  1983, MNRAS, 204, 1}

\bibitem{ms4a}{A. M. Burgess, \& R.W. Hunstead, 2006, AJ, 131, 100}

\bibitem{ms4b}{A. M. Burgess, \& R.W. Hunstead  2006, AJ, 131, 114}

\bibitem{2jy}{J. V.  Wall, \& J.A. Peacock  1985, MNRASm 216, 173}

\bibitem{morganti}{R. Morganti, et al. 1993, MNRAS, 263, 1023}

\bibitem{ackermann}{M. Ackerman,  et al.  2011, ApJ, 741, 30}

\bibitem{ghirlanda}{ G. Ghirlanda,  et al.  2010, MNRAS, 413, 852}

\bibitem{dermer}{C. D. Dermer  1995, ApJ, 446, L63}

\bibitem{torresi}{E. Torresi  2011,  Proceedings of Fermi \& Jansky: Our Evolving Understanding of AGN, St Michaels, MD, November 10-12, 2011, edited by R. Ojha, D. Thompson and C. Dermer, eConf  C1111101}

\bibitem{1275}{A.A. Abdo, et al.  2009, ApJ, 706, 275}

\bibitem{m87}{A.A. Abdo et al. 2009, 707, 55}

\bibitem{giulia}{G. Migliori, P. Grandi, E. Torresi, et al.  2011, A\&A, 533, 72}

\bibitem{paola2011}{P. Grandi  2011, Proceedings of the workshop "High Energy Phenomena in Relativistic Outflows III" (HEPRO III), Barcelona, June 27--July 1, 2011 (arXiv:1112.2505)}

\bibitem{dermerlott} {C. D. Dermer, $\&$ B. Lott B.  2011 Journal of Physics in press (arXiv:1110.3739)}


\end{thebibliography}
\end{document}